\documentclass{JHEP3}
\preprint{LU-TP 02-23\\
  hep-ph/0206195}

\usepackage{epsfig}
\usepackage{color}

\usepackage{graphics}
\usepackage{axodraw}
\usepackage{inputenc}
\usepackage{xspace}
\inputencoding{latin1}
 \renewcommand\email[1]{{\scriptsize\tt\href{mailto:#1}{#1}}}

\newcommand{\abar}{\ensuremath{\overline{\alpha}}}
\newcommand{\as}{\ensuremath{{\alpha}_{s}}}
\newcommand{\azero}{\ensuremath{{\alpha}_{0}}}

\renewcommand{\d}{\ensuremath{\mathrm{d}}}
\newcommand{\kT}{\ensuremath{k_{\perp}}}

\newcommand{\kTpot}[1]{\ensuremath{k_{\perp}^{#1}}}
\newcommand{\qbar}{\ensuremath{\overline{q}}}
\newcommand{\sigmahat}{\ensuremath{\hat{\sigma}}}
\newcommand{\smallx}{S\scalebox{0.8}{MALLX}\xspace}
\newcommand{\cascade}{C\scalebox{0.8}{ASCADE}\xspace}
\newcommand{\ldcmc}{\scalebox{0.8}{LDCMC}\xspace}
\def\mrm#1{\mathrm{#1}}
\def\sub#1{\ensuremath{_{\mrm{#1}}}}

\skip\footins = 1\bigskipamount plus 2pt minus 4pt                              

\title{\boldmath Gluon Distribution Functions\\ in the
  $\kT$-factorization Approach}

\author{G\"{o}sta Gustafson, Leif L\"onnblad and Gabriela Miu\\
  Dept.~of Theoretical Physics,
  S\"olvegatan 14A, S-223 62  Lund, Sweden\\
  E-mail: \email{Gosta.Gustafson@thep.lu.se}, \email{Leif.Lonnblad@thep.lu.se}
    and \email{Gabriela.Miu@thep.lu.se}}
  
  \abstract{At small $x$, the effects of finite transverse momenta of
    partons inside a hadron become increasingly important, especially
    in analyses of jets and heavy-quark production. These effects can
    be systematically accounted for in a formalism based on
    $\kT$-factorization and unintegrated distribution functions. We
    present results for the unintegrated distribution function, together
    with the corresponding integrated one, obtained within the
    framework of the Linked Dipole Chain model. Comparisons are made
    to results obtained within other approaches.}

\begin{document}
 
\sloppy

\section{Introduction}
\label{sect-intro}

In the description of a given cross section in deeply inelastic
lepton--hadron scattering (DIS), it is not enough to consider only the
leading order perturbative terms. Although \as\ may be small, each
power of \as\ may be accompanied by large logarithms due to the
large phase space available for additional gluon radiation.
It is, however, often possible to resum these emissions to all
orders in the leading logarithmic approximation (LLA).


The most familiar resummation strategy is based on
DGLAP~\cite{Gribov:1972ri,Lipatov:1975qm,Altarelli:1977zs,Dokshitzer:1977sg}
evolution, which resums large logarithms of the virtual photon
momentum transfer, $Q^2$. Within this formalism, the cross-section for
any given physics process is calculated using collinear factorization
of the form
\begin{equation}
  \sigma(x,Q^2) = \sigma_{0}(x,Q^2)\sum_a \int \frac{\d z}{z} \,
  C^{a}(z) \, f_{a}(\frac{x}{z},Q^2),
\label{eq:coll-fact} 
\end{equation}
i.e.\ as a convolution of coefficient functions $C^{a}$ and
parton densities $f_{a}(x,Q^2)$.

DGLAP evolution describes most experimental results\footnote{see e.g.\ 
  \cite{Anderson:2002cf} for a recent review.} from electron--proton
and proton--proton colliders. By using input parton densities which
are sufficiently singular when $x \rightarrow 0$, this formalism can
also account for the strong rise of $F_{2}$ for small $x$, as observed
at HERA.  However, there are problems with the description of
non-inclusive observables such as forward jet production in $ep$ and
heavy-quark production in $ep$ and $pp$ collisions.

In the region of very small $x$ (asymptotically large energies),
effects of finite transverse momenta of the partons may become more
and more important. The appropriate description in this region of
phase space is BFKL evolution~\cite{Kuraev:1977fs,Balitsky:1978ic}. The
cross-sections are calculated in the \kT-factorization approach of the
form
\begin{equation}
  \sigma(x,Q^2) = \int \frac{\d z}{z} \, \d^{2} \kT \,
  \sigmahat(z, Q^2, \kTpot{2}) \, \mathcal{F}(\frac{x}{z}, \kTpot{2}),
\label{eq:kt-fact} 
\end{equation}
i.e.\ as a convolution over the energy fraction $z$ and the transverse
momentum $\kT$ of the incoming parton of off-shell partonic
cross-sections $\sigmahat$, and $\kT$-unintegrated parton
densities\footnote{In this paper we will use $\mathcal{F}$ for the
  unintegrated parton distributions in general, and $\mathcal{G}$ for
  the unintegrated gluon distribution, treated as densities in
  $\log{1/x}$, i.e.\ $\mathcal{G}(x)=xG(x)$. For the integrated ones we
  will use the standard notation $f$ and $g$ respectively.},
$\mathcal{F}(x,\kTpot{2})$.  This corresponds to a resummation of
large logarithms of $1/x$. The BFKL evolution equation actually
predicts a strong power-like rise of $F_{2}$ at small $x$.

There exist a couple of models which take into account large
logarithms of both $Q^2$ and $1/x$ in DIS, reproducing both DGLAP and
BFKL in the relevant limits. One such model, valid for both small and
large $x$, has been developed by Ciafaloni, Catani, Fiorani and
Marchesini, and is known as the CCFM model
\cite{Ciafaloni:1988ur,Catani:1990yc}. The resulting unintegrated
distribution functions, $\mathcal{F}(x, \kTpot{2}, \qbar)$, depend on two
scales, the additional scale, $\qbar$, being a variable related to the
maximum angle allowed in the emission.

The Linked Dipole Chain model (LDC)
\cite{Andersson:1996ju,Andersson:1998bx} is a reformulation and
generalization of the CCFM model. Here, the unintegrated distribution
functions are essentially single-scale dependent quantities,
$\mathcal{F}(x, \kTpot{2})$. In this article we present results for the
integrated and unintegrated gluon distribution functions obtained
within the LDC formalism and make comparisons with the CCFM model and
with results from other formalisms.

This article is organized as follows. We start in section
\ref{sect-DIS} by giving a short introduction to the DGLAP and BFKL
formalisms for deeply inelastic $ep$ scattering, moving over to a
description of the CCFM model.  We end this section with a somewhat
more elaborate description of the Linked Dipole Chain model for DIS,
which is implemented in the Monte Carlo event generator
\ldcmc~\cite{Kharraziha:1998dn}.  This program can be used to describe
both structure functions and exclusive properties of the hadronic
final states.

Since the gluon distribution functions are not experimental observables they 
are not uniquely defined, but depend on the formalism used. We discuss
this problem, and some different approaches presented in the literature, 
in section \ref{sect-different}.

In section \ref{sect-results} we present our results for the
unintegrated and integrated gluon distribution functions, obtained in the
LDC formalism as implemented in \ldcmc.  These results are also
compared to those of other approaches, and we discuss how to make
relevant comparisons between the different formalisms.

We end this article with a summary in section \ref{sect-summary}.

%
%

\section{Deep Inelastic Scattering}
\label{sect-DIS}


Typically a deeply inelastic scattering event is represented by a fan
diagram, as the one shown in figure \ref{fig:fanDIS}. The (quasi-)real
emitted gluons, constituting the initial-state radiation, are labeled
$q_{i}$, while the virtual propagators are referred to as $k_{i}$. The
figure represents an exclusive final state, with the final-state
radiation explicitly marked as the dashed lines.  The final-state
emissions have to be defined so that they do not affect the
cross-section and give negligible recoils to the emitting partons. The
exact separation between initial- and final-state radiation depends,
however, on the formalism used. This problem will be further discussed
in section \ref{sect-different}.

We will here mostly discuss purely gluonic chains, which should give
the dominating contributions at small $x$. We will use different
approximations of the gluon splitting function
$P\sub{gg}(z)=\frac{1}{z}+\frac{1}{1-z}-2+z(1-z)$. Splitting a gluon
means that we have two new gluons, carrying fractions $z$ and $1-z$ of
the original gluon energy, which contribute to the gluon density,
while the contribution from the original one must be subtracted.

In analytical calculations the subtraction is achieved with the
so-called plus prescription for the pole at $z=1$ and the addition of a 
term proportional to $\delta(1-z)$ to the
splitting function. At asymptotically small $x$ the leading contribution
can be obtained by considering only the 
$1/z$ pole of the splitting function, thus effectively
only adding one gluon, neglecting the recoil for the emitting mother gluon.
In this way the problem with subtraction is avoided altogether. 
A third approach is to include both poles in the splitting
function
accounting for
the two produced gluons. The subtraction is then handled by a Sudakov
form factor, which multiplies each splitting and represents the
probability that the gluon to be split has not already been split
before. Here \emph{before} has to be defined by the ordering imposed
on the emissions (e.g.\ \kT-ordering in DGLAP and angular ordering in
CCFM). In an approximation where the non-singular terms in $P\sub{gg}$ 
(i.e.\ the terms $-2+z(1-z)$) are neglected, the Sudakov form factor, 
$\Delta\sub{S}$, is given by
\begin{equation}
  \label{eq:sudakov}
  \ln\Delta\sub{S} = - \int\abar\frac{dq_\perp^2}{q_\perp^2}\frac{dz}{1-z}
  \Theta\sub{order},
\end{equation}
where $\abar=3\as/\pi$. This definition is used in the CCFM approach.
Equivalently the Sudakov form factor can be
obtained by considering energy-momentum
conservation~\cite{Dokshitzer:1991wu} and be written
\begin{equation}
  \label{eq:yurisudakov}
  \ln\Delta\sub{S} = - \int\abar\frac{dq_\perp^2}{q_\perp^2}zdzP\sub{gg}(z)
  \Theta\sub{order}.
\end{equation}
Here the non-singular terms in $P\sub{gg}$ can be included without problems, 
and this method  is used in the \ldcmc generator.

To make the presentation more transparent, we will in the following 
discussion of the different
approaches only use the asymptotically small $x$ approximation, and
will return to the Sudakov form factor in section \ref{sect-results}.
There we will also consider the convolution of the perturbative
evolution with non-perturbative input parton densities, which will be
ignored in the remainder of this section.
 
\subsection{DGLAP and BFKL}
\label{subsect-DGLAP}

%
\FIGURE[t]{
\begin{picture}(170,230)(0,0)
\ArrowLine(10,15)(50,15)
\Text(20,25)[]{\large $proton$}
\Line(50,20)(100,20)
\Line(50,15)(100,15)
\Line(50,10)(100,10)
\Line(50,20)(80,40)
\GOval(50,15)(10,7)(0){1}
\Text(60,35)[]{\large $k_{0}$}
\Text(130,40)[]{\large $q_{1}$}
\Line(80,40)(120,40)
\Line(80,40)(95,70)
\Text(80,55)[]{\large $k_{1}$}
\Text(145,70)[]{\large $q_{2}$}
\DashLine(87.5,55)(110,55){2}
\Text(118,55)[]{\large $q'_{1}$}
\Text(90,85)[]{\large $k_{2}$}
\Text(155,100)[]{\large $q_{3}$}
\Line(95,70)(135,70)
\Line(95,70)(105,100)
\DashLine(100,70)(120,80){2}
\Line(105,100)(145,100)
\Line(105,100)(110,130)
\Line(110,130)(150,130)
\Line(110,130)(110,160)
\Line(110,160)(150,160)
\DashLine(120,160)(140,150){2}
\DashLine(130,155)(145,155){2}
\ArrowLine(20,220)(70,200)
\Text(166,160)[]{\large $q_{n+1}$}
\Text(100,145)[]{\large $k_{n}$}
\Text(20,210)[]{\large $lepton$}
\Text(100,185)[]{\large $q_{\gamma}$}
\ArrowLine(70,200)(120,220)
\Photon(70,200)(110,160){3}{5}
\end{picture}
  \caption{\label{fig:fanDIS} A fan diagram for a DIS event. 
    The quasi-real partons from the initial-state radiation are denoted
    $q_i$, and the virtual propagators $k_i$. The dashed lines denote
    final-state radiation.}}
In the DGLAP region, characterized by large $Q^2$ and limited $1/x$,
the dominating contributions come from $\kT$-ordered chains
which fulfill $Q^2\gg k_{\perp n}^2 \gg k_{\perp,n-1}^2 \gg \ldots$
and $k_{+ i} > k_{+,i+1}$. In the limit where also $x$ is small, so
that we can approximate the gluon splitting function with
$P_{gg}(z)\approx1/z$ (in the double leading log approximation
-- DLLA) we can write the unintegrated gluon distribution function for a
fixed coupling $\as$ on the form
\begin{eqnarray}
  \mathcal{G}(x,k_\perp^2) &\sim & \sum_n \int \prod^n \bar{\alpha} 
  \frac{dx_i}{x_i} \frac{d k_{\perp i}^2}{k_{\perp i}^2} 
  \theta(x_{i-1} -x_{i}) \theta(k_{\perp,i}^2 -k_{\perp,i-1}^2) \nonumber \\
  & \sim &  \sum_n \bar{\alpha}^n \frac{(\ln1/x)^n}{n!} 
  \frac{(\ln Q^2)^n}{n!} 
  \approx \exp (2\sqrt{\bar{\alpha} \ln Q^2 \ln 1/x}). \nonumber \\
  {\mathrm {where}} \,\,\,\, \bar{\alpha}\! & \equiv & \!\frac{3\alpha_s}{\pi} \,\,\,\,{\mathrm {and}} \,\,\,\, x_i \equiv k_{+i}/ P_{+,tot}.
  \label{eq:DGLAP}
\end{eqnarray}
In the case of a running coupling, $\abar(Q^2)\equiv\azero/\ln Q^2$, we get
the same exponential expression but with $\ln Q^2$ replaced by
$\ln(\ln Q^2)$ and \abar\ by \azero.

In the BFKL region of very small $x$ and limited $Q^2$, chains that
are not ordered in $\kT$ need to be accounted for, even though they
are suppressed. The resulting unintegrated distribution function
increases like a power at small $x$:
\begin{equation}
  \mathcal{G} \sim \frac{1}{x^\lambda} \, \mathrm{f}(\kT,x),
\end{equation}  
with the function $\mathrm{f}(\kT,x)$ describing a random walk in
$\ln(\kTpot{2}/\Lambda_{QCD}^2)$
\cite{Levin:1990gg,Bartels:1993du,Forshaw:1994es,Andersson:1998bx}.
Such a power-like behavior is in approximate agreement with HERA data,
with $\lambda \sim 0.3$. We note, however, that a corresponding
increase is also obtained from NLO DGLAP evolution.

Both the DGLAP and BFKL evolution were developed to describe inclusive
quantities such as $F_2$, but they can be interpreted as an explicit
summation of initial-state brems\-strahlung (ISB) of quasi-real
partons, and can thus be used to describe exclusive multi-parton 
final states.  To do this we must also include the final-state
brems\-strahlung (FSB) from the ISB partons within kinematic regions
allowed by the colour coherence constraint.  This final-state
radiation should also be emitted in such a way that it gives
negligible recoils, and that it does not affect the total cross
section. The separation between ISB and FSB depends upon the formalism 
used, and if more partons are treated as
initial-state radiation we get a larger number of contributing chains,
which is compensated by smaller weights for each chain, and with
correspondingly reduced kinematic regions for final-state emissions.

\subsection{CCFM}
\label{subsect-CCFM}

The particular calculation scheme adopted by Ciafaloni, Catani,
Fiorani and Marchesini has resulted in the well-known CCFM model
\cite{Ciafaloni:1988ur,Catani:1990yc}.  This model, that has been
developed assuming purely gluonic chains, provides a description not
only of the structure function evolution in DIS, accurate at the
leading-log level, but also of final-state partons. Colour coherence
implies that the initial-state emissions are ordered in angle (or
equivalently in rapidity). According to the definition of the
separation between initial- and final-state radiation, they are also
ordered in the positive (along the incoming proton) light-cone
momentum $q_+$. All other kinematically allowed emissions (symbolized
by the $q'_1$ emission in figure \ref{fig:fanDIS}) are defined as
final-state emissions.

The CCFM model is based on the $\kT$-factorization formalism, with the
unintegrated distribution function in the small-$x$ limit given by:
\begin{eqnarray}
  \label{fCCFM}
  \mathcal{G}(x,k_\perp^2,\qbar) &\sim& \sum_n \int \prod^n \bar{\alpha} 
  \frac{dz_i}{z_i} \frac{d^2 q_{\perp i}}{\pi q_{\perp i}^2} 
  \Delta_{ne} (z_i, k_{\perp i}^2, \qbar_i)\times\\
  & & \delta(x-\Pi z_i)
  \theta(\qbar_i - \qbar_{i-1} z_{i-1})
  \delta(k_\perp^2-k_{\perp n}^2) \theta(\qbar-\qbar_n z_n)\nonumber.
\end{eqnarray}
The notation is that of figure \ref{fig:fanDIS}, i.e.\ $q_{\perp,i}$
and $k_{\perp,i}$ are the transverse momenta of the real and virtual
partons, respectively. The splitting parameter $z$ is defined as
$z_i=k_{+,i}/k_{+,i-1}$, and the so-called rescaled transverse
momentum $\qbar_i$ is defined by $\qbar_i \equiv q_{\perp i}/(1-z_i)$.
The interval for the $z$-variables is between 0 and 1, which
guarantees ordering in $q_+$, and the angular ordering condition is
satisfied by the constraint
\begin{equation}
 \label{ao} 
 \qbar_i > \qbar_{i-1} z_ {i-1},
\end{equation}
explicitely written out in eq.~(\ref{fCCFM}).  Moreover, we note the
occurrence of the so-called non-eikonal form factor $\Delta_{ne} (z,
k_{\perp}^2, \qbar)$ defined in ref.~\cite{Catani:1990yc}. The
distribution function, $\mathcal{G}$, depends on two separate scales,
the transverse momentum, \kT, of the interacting gluon, and \qbar,
which determines an angle beyond which there is no (quasi-) real
parton in the chain of initial-state radiation. In the rest frame of
the incoming proton, this limiting angle corresponds to a rapidity
given by (if counted negative in the direction of the probe)
\begin{equation}
  \label{eq:ylim}
  y\sub{lim}=\ln\left(x\frac{m_p}{\qbar}\right)
\end{equation}
In the original formulation there was also the so-called consistency
constraint
\begin{equation}
 k_{\perp i}^2 > z_i q_{\perp i}^2,
\label{cc}
\end{equation}
which was needed to guarantee that the virtuality $k^2$ is well
approximated by $-\kT^2$. This constraint has a non-leading effect,
and has been disregarded in some analyses \cite{Kwiecinski:1996td}.

The CCFM evolution has been implemented in two hadron-level event
generators, \smallx\cite{Marchesini:1991zy,Marchesini:1992jw} and
\cascade\cite{Jung:2001hx}, both maintained by Hannes Jung. These 
programs reproduce HERA data on $F_{2}$ well for small $x$, where
purely gluonic chains should give the dominating contribution. For
larger $x$, we expect a large contribution from valence quarks;
such chains are not easily accounted for in the CCFM formalism and 
are not included in the programs.

Both programs are able to reproduce a wide range of final-state
observables. We note, however, that there is a large sensitivity to
non-leading corrections. In particular, the description of forward-jet
production at HERA turns out to be very sensitive to the non-singular
terms in the gluon splitting function. In the original CCFM
formulation these terms were left out, and without them the
forward-jet rates are well described. If, however, the non-singular
terms are included, which would be the most natural option, the jet
rates come out approximately a factor two below the data
\cite{Anderson:2002cf}.

\subsection{The Linked Dipole Chain Model}
\label{sect-LDC}






The Linked Dipole Chain model (LDC) is based on the CCFM model, and
agrees with CCFM to leading double log accuracy. Also LDC is
formulated in terms of $\kT$-factorization and unintegrated
distribution functions.  In LDC the ISB definition has been modified,
resulting in a more simple description, with the unintegrated
distribution functions being (essentially) dependent on only one scale
and allowing for some sub-leading corrections to be introduced in a
rather straight-forward manner.

In LDC more gluons are treated as final-state radiation. The remaining
initial-state gluons are ordered both in $q_+$ {\it and} $q_-$
(which implies that they are also ordered in angle or rapidity $y$) with
$q_{\perp i}$ satisfying 
\begin{equation}
  q_{\perp i} > \min(k_{\perp i},k_{\perp,i-1}).
  \label{eq:ldccut}
\end{equation}
This redefinition of the ISB--FSB separation implies that one single
chain in the LDC model corresponds to a set of CCFM chains.  It turns
out that when one considers the contributions from all chains of this
set, with their corresponding non-eikonal form factors, they add up to
one \cite{Andersson:1996ju}. Thus, the non-eikonal form factors do not
appear explicitly in LDC, resulting in a simpler form for the
unintegrated distribution function
\begin{equation}
  \mathcal{G}(x,k_\perp^2) \sim \sum_n \int \prod^n \bar{\alpha} 
  \frac{dz_i}{z_i} \frac{d^2 q_{\perp i}}{\pi q_{\perp i}^2} 
  \theta(q_{+,i-1} -q_{+ i}) \theta(q_{- i} -q_{-,i-1}) 
  \delta(x-\Pi z_i) \delta(\ln k_\perp^2 - \ln k_{\perp n}^2).
  \label{eq:fq}
\end{equation}

The notation in the above and what will follow refers to that of
figure \ref{fig:typ-chain}. Here, a typical DIS event is shown
together with the corresponding phase space available in the
$\gamma$-p rest frame, where the rapidity, $y$, and the transverse
momentum, $q_{\perp}$, of any final-state parton are limited by a
triangular region in the ($y,\ln q_{\perp}^2$)-plane. The proton
direction is towards the right end of the triangle, and the photon
direction is towards the left. The real emitted (ISB) gluons are
represented by points in this diagram.  The virtual propagators do not
have well defined rapidities, and are represented by horizontal lines,
the left and right ends of which have the coordinates $(\ln [k_{+
  i}/k_{\perp i}],\ln k_{\perp i}^2$) and ($\ln [-k_{\perp i}/k_{-
  i}],\ln k_{\perp i}^2$) respectively. The phase space available for
FSB is given by the area below the horizontal lines (including the
folds that stick out of the main triangle).
%
%
\FIGURE[t]{\scalebox{0.9} {\mbox{
\begin{picture}(140,230)(0,-50)
  \Line(10,15)(50,15)
  \Line(50,20)(80,20)
  \Line(50,15)(80,15)
  \Line(50,10)(80,10)
  \Line(50,15)(60,40)
  \GOval(50,15)(10,7)(0){1}
  \Line(60,40)(90,40)
  \Line(60,40)(70,70)
  \Line(70,70)(105,70)
  \Line(70,70)(75,100)
  \Line(75,100)(110,100)
  \Line(75,100)(80,130)
  \Line(80,130)(110,130)
  \Line(80,130)(85,160)
  \Line(85,160)(120,160)
  \Line(85,160)(85,190)
  \Line(85,190)(120,190)
  \Photon(50,210)(85,190){3}{4}
  \Line(85,190)(120,190)
  \Line(20,220)(50,210)
  \Line(50,210)(80,220)
  \Text(47,34)[]{\large $k_{0}$}
  \Text(55,55)[]{\large $k_{1}$}
  \Text(65,85)[]{\large $k_{2}$}
  \Text(70,115)[]{\large $k_{3}$}
  \Text(75,145)[]{\large $k_{4}$}
  \Text(75,173)[]{\large $k_{5}$}
  \Text(100,40)[]{\large $q_{1}$}
  \Text(115,70)[]{\large $q_{2}$}
  \Text(120,100)[]{\large $q_{3}$}
  \Text(125,130)[]{\large $q_{4}$}
  \Text(130,160)[]{\large $q_{5}$}
  \Text(130,190)[]{\large $q_{6}$}
\end{picture}}}
\scalebox{0.8} {\mbox{
\begin{picture}(340,280)(0,-20)
  \Line(40,20)(300,20)
  \Line(40,20)(170,280)
  \Line(170,280)(300,20)
  \Line(80,100)(120,20)
  \Text(60,85)[]{$q_{6}$}
  \Vertex(70,80){2}
  \Text(110,70)[]{$k_{5}$}
  \Vertex(125,80){2}
  \Text(125,90)[]{$q_{5}$}
  \Line(90,80)(125,80)
  \Line(125,80)(130,70)
  \Line(130,70)(145,70)
  \Text(140,60)[]{$k_{4}$}
  \Line(145,70)(160,100)
  \Line(160,100)(190,100)
  \Vertex(160,100){2}
  \Text(160,110)[]{$q_{4}$}
  \Text(175,90)[]{$k_{3}$}
  \Vertex(190,100){2}
  \Text(190,110)[]{$q_{3}$}
  \Line(190,100)(200,80)
  \Line(200,80)(230,80)
  \Text(215,70)[]{$k_{2}$}
  \Line(230,80)(250,40)
  \Vertex(230,80){2}
  \Text(230,90)[]{$q_{2}$}
  \Line(250,40)(290,40)
  \Text(270,30)[]{$k_{1}$}
  \Vertex(290,40){2}
  \Text(300,50)[]{$q_{1}$}
  \LongArrow(250,180)(250,230)
  \LongArrow(250,180)(300,180)
  \Text(250,240)[]{$\ln \kTpot{2}$}
  \Text(310,180)[]{$y$}
  \DashLine(125,80)(125,20){2}
  \Line(125,80)(135,10)
  \Line(125,20)(135,10)
  \DashLine(160,100)(160,20){2}
  \Line(160,100)(180,0)
  \Line(180,0)(160,20)
  \DashLine(190,100)(190,20){2}
  \Line(190,100)(210,0)
  \Line(210,0)(190,20)
  \DashLine(230,80)(230,20){2}
  \Line(230,80)(240,10)
  \Line(240,10)(230,20)
  \DashLine(40,-10)(40,20){2}
  \DashLine(120,-10)(120,20){2}
  \DashLine(300,-10)(300,20){2}
  \LongArrow(40,-5)(120,-5)
  \LongArrow(120,-5)(40,-5)
  \LongArrow(120,-5)(300,-5)
  \LongArrow(300,-5)(120,-5)
  \Text(80,-15)[]{$\ln Q^2$}
  \Text(210,-15)[]{$\ln 1/x$}
  \LongArrow(220,130)(260,150)
  \Text(280,150)[]{$\ln q_+$}
  \LongArrow(120,130)(80,150)
  \Text(60,150)[]{$\ln q_-$}
  \DashLine(15,100)(80,100){2}
  \DashLine(15,20)(40,20){2}
  \LongArrow(20,20)(20,100)
  \LongArrow(20,100)(20,20)
  \Text(0,60)[]{$\ln Q^2$}
\end{picture}}}

  \caption{\label{fig:typ-chain} The initial-state emissions $q_i$
    in the $(y,\kappa=ln(k_{\perp}^2))$-plane. Final-state radiation
    is allowed in the region below the horizontal lines. The height of
    the horizontal lines determine $\ln k_{\perp i}^2$. The light-cone
    momenta $k_{+i}$ and $k_{-i}$ can be read off as described in the
    text.}}
  
The ordering of the CCFM evolution in $q_+$ but not in $q_-$ means
that this formalism is not left--right symmetric. In contrast the LDC
formulation is completely symmetric, which implies that the chain in
figure \ref{fig:typ-chain} can be thought of as evolved from either
the photon or the proton end. (Thus, the LDC formalism automatically
takes into account contributions from ``resolved photons''.)

Returning to eq.~(\ref{eq:fq}), we note that it can equally well be
expressed in terms of the virtual propagator momenta. Due to the
condition in eq.~(\ref{eq:ldccut}) we have $q_{\perp i}^2 \approx
\max(k_{\perp i}^2,k_{\perp,i-1}^2)$ and, suppressing the $\theta$-
and $\delta$-functions, we obtain
\begin{equation}
  \mathcal{G} \sim \sum \int \prod \bar{\alpha} \frac{dz_i}{z_i} 
  \frac{d k_{\perp i}^2}{\max(k_{\perp i}^2,k_{\perp,i-1}^2)}.
  \label{eq:fk}
\end{equation}
In particular we note that this implies that, for a ``step up'' or a
``step down'' in $\kT$, the following weights result:
\begin{eqnarray} 
  \frac{d^2 q_{\perp i}}{q_{\perp i}^2} \approx \frac{d^2 k_{\perp i}}
  {k_{\perp i}^2},
  \,\,\,\, k_{\perp i} > k_{\perp,i-1} \,\,\,\,\,{\mathrm {and}}
  \label{eq:step-up}\\
  \frac{d^2 q_{\perp i}}{q_{\perp i}^2} \approx \frac{d^2 k_{\perp i}}
  {k_{\perp i}^2} 
  \cdot \frac{ k_{\perp i}^2}{k_{\perp,i-1}^2},\,\,\,\,\,
  k_{\perp i} < k_{\perp,i-1}.
  \label{eq:step-down}
\end{eqnarray}
Thus, for a step down there occurs an additional suppression factor $
k_{\perp i}^2/k_{\perp,i-1}^2$.

The relation between the integrated and the unintegrated distribution
functions are symbolically written as $xf(x,Q^2) \sim \int \frac{d
  k_{\perp}^2}{k_{\perp}^2} \mathcal{F}(x,k_{\perp}^2)$. The exact
relationship is somewhat dependent on the evolution scheme used for
$\mathcal{F}(x,k_{\perp}^2)$. For LDC we have
\begin{equation}
  xf(x,Q^2) = \int^{Q^2} \frac{d k_{\perp}^2}{k_{\perp}^2}
  \mathcal{F}(x,k_{\perp}^2) + \int_{Q^2} \frac{d k_{\perp}^2}{k_{\perp}^2}
  \mathcal{F}(x \frac{k_\perp^2}{Q^2},k_{\perp}^2) 
  \frac{Q^2}{k_{\perp}^2},
  \label{eq:relation-new}
\end{equation}
assuming a single-scale dependent unintegrated distribution function.

The first term of eq.~(\ref{eq:relation-new}) corresponds to chains
whose struck parton is less virtual than the probe, $\kTpot{2} < Q^2$.
The second term contains the suppressed contributions that come from
chains whose struck parton has $\kTpot{2} > Q^2$; the suppression
factor $Q^2 / k_{\perp}^2$ is analogous to the one occurring in
eq.~(\ref{eq:step-down}). Note also that the $x$-argument in the
unintegrated distribution function of the second term has been rescaled 
by the factor $k_{\perp}^2 / Q^2$.

Below we list some properties of the LDC model: 
\begin{itemize}
\item The natural scale in the running coupling, $\as$, is $q_{\perp
    i}^2$, which coincides with $\max(k_{\perp i}^2, k_{\perp,i-1}^2)$.
\item Non-leading effects such as those coming from quark-initiated
  chains, from the non-singular terms in the splitting functions and
  from energy-momentum conservation can be included in a
  straight-forward manner. It is also possible to correct the two
  emissions connected with the highest-virtuality link with the full
  $2\rightarrow2$ matrix element, thus improving the result. Finally,
  for the ISB emission closest to the photon end, the full off-shell
  $\mathcal{O}(\alpha\as)$ matrix element is used.
\item The fact that fewer gluons are considered as initial-state
  radiation implies that typical $z$-values are smaller, thus
  resulting in smaller sub-leading corrections.
\item The formalism is greatly simplified by the fact that the
  non-eikonal form factors do not appear explicitly in the model.
\item The LDC formalism is fully left-right symmetric, in the sense
  that the result is independent of whether the evolution starts at
  the probe (photon) or target (proton) end. Hence it can
  easily be generalized to the case when the photon is replaced by a
  hadron. As a result, the formalism can be applied to the study of
  jet production in hadronic collisions \cite{Gustafson:1999kh}.
\item Besides giving an inclusive description of the events, the
  result in eq.~(\ref{eq:fk}) can also be interpreted as the
  production probability for an exclusive final state.
\end{itemize}
These qualities make LDC particularly suitable for implementation in
an event generator. One such LDC-based Monte Carlo event generator, \ldcmc, has
been developed by Kharraziha and L\"{o}nnblad~\cite{Kharraziha:1998dn}.

The \ldcmc has been shown to be able to reproduce available $F_2$ data
very well (see further section \ref{sect-results}), not only the 
small-$x$ data from of HERA but also data at
higher $x$ from fixed target experiments. This success relies on the
fact that quark-initiated chains are easily incorporated in the
evolution.

It should be mentioned that, just like the CCFM-based \cascade,
\ldcmc has problems describing the H1 forward-jet data. Again
the data can be reproduced only when the non-singular terms of the
gluon splitting function are omitted. The more physical case of
including the full splitting function predicts a distribution which is
approximately a factor 2 below the data.

In section \ref{sect-results} we present results for the integrated
and unintegrated gluon distribution functions obtained within the LDC
formalism using the \ldcmc simulation program. Before comparing these
results with those of other approaches, we discuss in the following
section the relationship between the integrated and unintegrated
distribution functions in different formalisms.

\section{Discussion on Different Distribution Function Definitions}
\label{sect-different}

In contrast to the structure function $F_2$, the unintegrated parton
distributions are not experimental observables, and depend on the
specific scheme in which they are defined. Not only do they depend on
the approximations used in the description of the evolution, it has
also been argued that the unintegrated distributions are not
gauge-invariant objects and there is a dependence on the gauge choice
used for the off-shell matrix elements they are convoluted
with\footnote{See e.g.\ discussions in \cite{Anderson:2002cf} and
  \cite{DokshitzerLundSmallx2}}.

For asymptotically large values of $Q^2$ or $1/x$ it is possible to
define unintegrated parton distribution functions,
$\mathcal{F}(x,\kT^2)$, which depend on a single scale, $\kT^2$,
specifying the transverse momentum (or virtuality) of the interacting
parton. In the DGLAP region the $\kT$-ordering of the links implies
that we can define $\mathcal{F}(x,\kT^2)$ as the derivative of the
integrated distribution functions, $xf(x,Q^2)$, and thus
\begin{equation}
xf(x,Q^2) = \int^{Q^2} \frac{d \kT^2}{\kT^2} \mathcal{F}(x,\kT^2)
\label{dglap-non-int}
\end{equation}
Also in the BFKL equation the distribution functions are determined by a
single scale. However, since the chains are not ordered in $\kT$, the
integrated distribution function, $f(x,Q^2)$, also obtains contributions
from $\kT$-values larger than $Q^2$, and therefore the simple relation
in eq.~(\ref{dglap-non-int}) is not satisfied.

As discussed in section \ref{subsect-CCFM} the angular ordering in the
CCFM model implies that the unintegrated distribution functions depend
on two scales, $\kT$ and $\qbar$, where \qbar\ specifies the limiting
angle in eq.~(\ref{eq:ylim}). (Note that the strong ordering in $\kT$ in
DGLAP, or in $x$ in BFKL, also automatically implies an ordering in
angle, which makes this second scale redundant.)  In the CCFM
formalism the gluon distribution function depends very strongly on the
scale $\qbar$, when $\qbar$ is in the neighbourhood of $\kT$.  The
angular constraint is expressed by the last $\theta$-function in
eq.~(\ref{fCCFM}), which guarantees the relation
\begin{equation}
  \qbar > \qbar_n z_n, \,\,\,\mathrm{where}\,\,\,\qbar_n \equiv 
  \frac{q_{\perp n}}{1 - z_n}
\label{angularlimit}
\end{equation}
This implies an upper limit for $z_n$ of the last emission given by
\begin{equation}
  z_n < z\sub{lim} =  \frac{\qbar}{\qbar + q_{\perp}}
\label{z-limit}
\end{equation}
From this result we can see that if the scale $\qbar$ is chosen in the
neighborhood of $\kT$, then a large fraction of the possible chains
are cut away.  To realize this we make two observations:
\begin{itemize}
\item[(i)] For $\qbar = \kT$ we find $z\sub{lim} \approx 0.5$ when the
  last link is a step up in $\kT$ (in which case $q_\perp \approx
  \kT$), and $z\sub{lim} < 0.5$ when the last link is a step down (and
  $q_\perp > \kT$).
\item[(ii)] In the CCFM model the $1/z$ pole in the splitting function
  is screened by the non-eikonal form factor, and the $z$-distribution
  obtained in the \smallx and \cascade MCs therefore does not peak at
  small $z$-values, but has a maximum around
  $z=0.5$~\cite{Anderson:2002cf}.
\end{itemize}
This implies that for $\qbar = \kT$ the constraint in
eq.~(\ref{z-limit}) will exclude a large fraction of the possible
chains.  Furthermore, the fact that for $\qbar = \kT$, $z\sub{lim}$
approximately coincides with the maximum in the $z$-distribution
implies that increasing (decreasing) $\qbar$ in the neighborhood of
$\kT$ includes (excludes) a significant set of chains. Consequently,
for fixed $\kT$, the structure functions depend strongly on $\qbar$ in
this region.

The relevant values for $\qbar$ in a hard sub-collision should, however, 
be significantly larger than $\kT$. If the limiting angle is given by 
the final state parton in the hard collision, then it is easy to show that 
\begin{equation}
  \qbar^2 =  \frac{\hat{t} \hat{s}}{\hat{u}}
\label{qbarhard}
\end{equation}
where $\hat{t}$, $\hat{s}$, and $\hat{u}$ are the Mandelstam variables
for the sub-collision. Thus if $\hat{s}$ is large compared to
$\hat{t}$ we find $\qbar^2 \approx - \hat{t}$. For a very hard
collision we find $\qbar^2$ of the same order as $\hat{s}$. We note
that choosing $\qbar^2=\hat{s}$ corresponds to a limiting angle equal
to 90$^\circ$ in the rest frame of the hard sub-collision. For a
typical hard sub-collision we may thus have $\qbar$ substantially
larger than $\kT$, and we will return to this question in the
following section.

Many of the gluons which make up the initial-radiation chain in the
CCFM model, are treated as final-state radiation in the LDC formalism.
Therefore typical $z$-values are smaller in the LDC model, 
and most of the problem of
angular ordering is postponed to the treatment of the final-state
radiation. To leading order in $\ln 1/x$ the result is determined by
the $1/z$ pole, and the unintegrated distribution function in LDC depends
on only a single scale, $\kT^2$. As discussed in section
\ref{sect-DIS}, sub-leading effects due to the $1/(1-z)$ pole or the
non-singular terms in the splitting function are included with Sudakov
form factors, which do depend on the angular region allowed for
radiation. Therefore also in LDC the unintegrated distribution
functions depend on the scale $\qbar$ defined above, although as we
will see below, the dependence is very much weaker for this model.

Many schemes are presented in the literature to treat unintegrated
parton distributions. Besides with the CCFM formalism in the \cascade
and \smallx MCs, which in the following will be referred to as JS
(Jung and Salam) \cite{Jung:2000hk}, we compare in the following
section our results also with the formalisms presented by Kwiecinski,
Martin, and Stasto (KMS) \cite{Kwiecinski:1997ee} and by Kimber,
Martin, and Ryskin (KMR) \cite{Kimber:2001sc}.  In KMS a term
describing leading order DGLAP evolution is added to the BFKL
equation. The parton distribution is described by a single scale,
$\kT$, and is assumed to satisfy the relation in
eq.~(\ref{dglap-non-int}). In KMR two-scale parton distributions are
extracted from the same unified DGLAP-BFKL evolution equation, but as
we discuss in more detail in section \ref{subsect-unintegrated}, the
dependence on $\qbar$ for fixed $\kT$ is rather weak. Finally we will
compare to a simple derivative, according to
eq.~(\ref{dglap-non-int}), of the integrated gluon density from the
GRV98 \cite{Gluck:1998xa} parameterization, referred to as dGRV in the
following.

\section{Results}
\label{sect-results}

In this section we discuss some results obtained from \ldcmc. To
illuminate the effect of the different contributions we study the
following three different versions:

\begin{enumerate}
\item {\em Standard}: Including non-leading contributions from quarks
  and non-singular terms in the splitting functions.
\item {\em Gluonic}: Including non-singular terms in the splitting
  functions, but no quark links in the evolution.
\item {\em Leading}: Also purely gluonic chains, but with only the
  singular terms in the gluon splitting function.
\end{enumerate}
To get realistic results from these three versions we now also need to
consider the convolution of non-perturbative input parton densities.
These are not {\`a} priori known, but need to be parameterized in some
way.  We will use the same parameterization as in
\cite{Kharraziha:1998dn} given by
\begin{equation}
xf_i(x,k_{\perp0}^2) = A_i x^{a_i} (1-x)^{b_i},
\label{eq:fit-form}
\end{equation}
where $i=d_v,u_v,g$ and $s$ for the d-valence, u-valence, gluon and
sea-quark densities respectively (where the sea flavour densities are
assumed to be $f_{\bar{d}}=f_{\bar{u}}=2f_{\bar{s}}$). The parameters
$A_i,a_i,b_i$ and the perturbative cutoff, $k_{\perp0}$, are then
fitted to reproduce the measured data on $F_2$. There are some sum
rules which fix the relationship between some of the parameters. The
$A_{d_v}$ and $A_{u_v}$ are fixed by flavour conservation and $A_s$ is
fixed by momentum conservation. The fits to $F_2$ do not constrain the
remaining parameters very strongly, so we have fixed the $b$
parameters to $3$ in the valence densities and to $4$ in the sea and
gluon densities. To check the sensitivity to the $b$ parameter we have
an additional fit for the \textit{gluonic} case with $b=7$ called
\textit{gluonic-2}. In table \ref{tab:param} we present the result of
the fits. Note that in the case of the \textit{gluonic} and
\textit{leading} versions, only the gluon input density is considered.

\TABLE[t]{
  \begin{tabular}{|l||r|r|r||r|r||r|r||r|r||r||r||l|}
    \hline
    fit & $A_g$ & $a_g$ & $b_g$ &
    $a_d$ & $b_d$ & $a_u$ & $b_u$ & $a_s$ & $b_s$ & $k_{\perp 0}$ &
    $\int xg(x)$ & $\chi^2/\mbox{d.o.f.}$ \\
    \hline
    \textit{standard} & 1.86 & \textbf{0} & \textbf{4} &
    1.78 & \textbf{3} & 0.57 & \textbf{3} &
    \textbf{0} & \textbf{4} & 0.99 & \textit{0.37} & 694/625 \\
    \textit{gluonic} & 2.71 & \textbf{0} & \textbf{4}
    & & & & & & & 1.80 & \textit{0.54} & 193/86 \\
    \textit{gluonic-2} & 3.11 & \textbf{0} & \textbf{7}
    & & & & & & & 2.17 & \textit{0.39} & 125/86 \\
    \textit{leading} & 2.34 & \textbf{0} & \textbf{4}
    & & & & & & & 1.95 & \textit{0.47} & 126/86 \\
    \hline
  \end{tabular}
  \caption{The result of the fit of the parameters for the
    input parton densities. The \textit{standard} version has been
    fitted to data from H1 \cite{Aid:1996au},
    ZEUS \cite{Derrick:1996hn}, NMC \cite{Arneodo:1995cq} and
    E665 \cite{Adams:1996gu} in the region $x<0.3$, $Q^2>1.5$~GeV$^2$,
    while the \textit{gluonic} and \textit{leading} have been fitted
    to H1 data only, in the region $x>0.013$ and $Q^2>3.5$~GeV$^2$.
    The last two columns give the resulting fraction of the proton
    momentum carried by the gluons and the $\chi^2$ over the number of
    fitted data points, respectively. Parameters in bold face have not
    been fitted.}
  \label{tab:param}}

\FIGURE[t]{
  \epsfig{figure=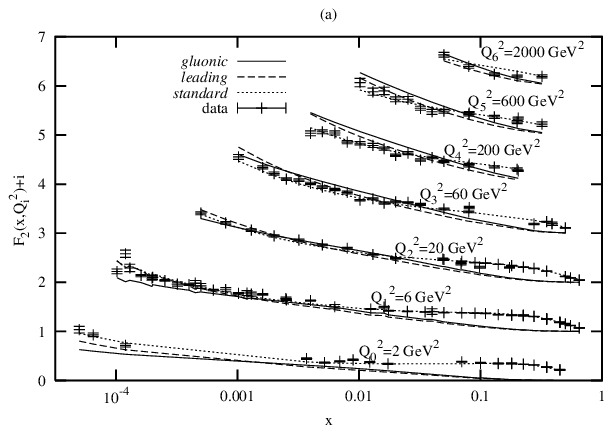,width=10cm}
  \epsfig{figure=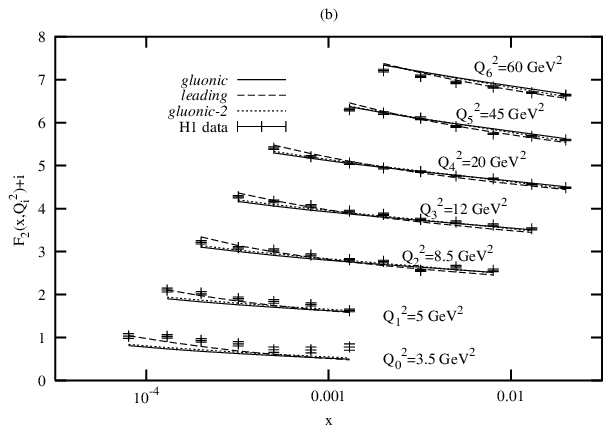,width=10cm}
  \caption{The description of $F_2$ data as a
    function of $x$ for the different fits presented in this paper.
    To separate the results for different $Q^2$, what is shown is
    $F_2+i$ for $Q_i^2$. In (a) both small and large $x$ data from H1,
    ZEUS \cite{Derrick:1996hn}, NMC \cite{Arneodo:1995cq} and
    E665 \cite{Adams:1996gu} are included. The full line is
    \textit{gluonic}, dashed is \textit{leading} and the dotted line
    is \textit{standard}. In (b) the versions with only gluonic chains
    are compared to small-$x$ H1 data \cite{Aid:1996au}.  As in (a),
    full line is the \textit{gluonic} case, dashed is
    \textit{leading}, whereas dotted is \textit{gluonic-2} (with
    $b_g$=7).}
  \label{fig:f2}}
  
In figure \ref{fig:f2} we show the resulting reproduction of $F_2$
data. Clearly, all versions give a satisfactory fit to the data.  The
\textit{standard} version gives an excellent description for all
values of $x$. The \textit{gluonic} and \textit{leading} versions are
naturally unable to fit the large-$x$ data, as the inclusion of
quark-initiated chains is essential for a description of large
$x$-values.  These versions are therefore only fitted to the small-$x$
H1 data, and these fits are shown separately in figure \ref{fig:f2}b.
We note that \textit{gluonic-2} gives a better fit than
\textit{gluonic}, but we will anyway in the following concentrate on
the latter, to make the comparison with the JS results (which also
uses $b_g=4$) more informative.

It should be noted that the version of \ldcmc used here has not been
released yet, and differs somewhat from the one described in
\cite{Kharraziha:1998dn}. The main difference is the handling of the
Sudakov form factors, which were not quite correct in the original
version. Also the full off-shell $\gamma^*g^*\rightarrow q\bar{q}$
matrix element is now included\footnote{Note, however, that the matrix
  element in \cite{Baier:1981kx} is used, rather than the one from
  \cite{Catani:1991eg} used in \smallx and \cascade}. These, and other
minor changes, do not give a big effect on the results.

In the following subsections we discuss the corresponding results for
the integrated and the unintegrated gluon distribution functions. As
we will see, the results are very similar for the \textit{standard}
and the \textit{gluonic} version. The differences in the cascade can
here be compensated by the adjustment of the input distribution
functions and the perturbative cutoff, $k_{\perp 0}$. For the
\textit{leading} version, without the non-singular terms in the
splitting function, we find, however, that the resulting differences
are small but not negligible.

\subsection{Results for the Integrated Gluon Distribution Function}
\label{subsect-integrated}

We start by studying the results obtained for the integrated gluon
distribution function. As mentioned in section \ref{sect-LDC} (cf.\ 
eq.~(\ref{eq:relation-new})), in the LDC model the relation between
the integrated and the unintegrated gluon distribution functions is as
follows:
\begin{equation}
  xg(x,Q^2) = \int^{Q^2}_{k_{\perp 0}^2} \frac{\d \kTpot{2}}{\kTpot{2}}
  \mathcal{G}(x,\kTpot{2}, Q) + \int^{Q^2/x}_{Q^2}
  \frac{\d \kTpot{2}}{\kTpot{2}}
  \mathcal{G}(x \frac{\kTpot{2}}{Q^2},\kTpot{2}, Q)
  \frac{Q^2}{\kTpot{2}} + xg_{0}(x,Q_0^2) \times \Delta\sub{S}.
  \label{eq:int-terms} 
\end{equation}
Thus, the integrated distribution function receives contributions from
three different terms. The first term corresponds to struck gluons of
transverse momenta below the virtuality of the probe, $\kTpot{2} <
Q^2$ (the full line in the triangular phase space of figure
\ref{fig:hard-probe} is an example of such a chain), while the second
term originates from chains whose struck gluons have $\kTpot{2} > Q^2$
(e.g.\ the long-dashed line in figure \ref{fig:hard-probe}).  Finally,
the third term is the contribution from the input distribution
function, corresponding to the case when no evolution has taken place.

As discussed in section \ref{sect-LDC}, the distribution function
$\mathcal{G}(x,\kTpot{2}, Q)$ in the LDC model depends only very
weakly upon the scale $Q$. This dependence is due to the Sudakov
form factor corresponding mainly to the $1/(1-z)$ term in the
splitting function. Although this dependence is weak and can
essentially be neglected in the first two terms in
eq.~(\ref{eq:int-terms}), it has a larger effect on the last term,
where it is explicitely included as a multiplicative factor. This term
dominates at large $x$, and the effect of the form factor is a
suppression for larger values of $Q^2$.
\FIGURE[t]{\scalebox{0.85}{\mbox{
\begin{picture}(340,280)(0,0)
  \Line(50,40)(290,40)
  \Line(50,40)(170,280)
  \Line(170,280)(290,40)
  \Line(90,120)(130,40)
  \DashLine(110,160)(170,40){2}
  \Line(100,100)(130,100)
  \Line(130,100)(140,80)
  \Line(140,80)(180,80)
  \Line(180,80)(190,100)
  \Line(190,100)(225,100)
  \Line(225,100)(245,60)
  \Line(245,60)(280,60)
  \DashLine(110,160)(140,160){6}
  \DashLine(140,160)(145,150){6}
  \DashLine(145,150)(185,150){6}
  \DashLine(185,150)(190,140){6}
  \DashLine(190,140)(210,140){6}
  \DashLine(210,140)(230,110){6}
  \DashLine(230,110)(255,110){6}
  \DashLine(30,120)(90,120){2}
  \DashLine(30,40)(50,40){2}
  \LongArrow(35,42)(35,118)
  \LongArrow(35,118)(35,42)
  \Text(16,80)[]{\large $\ln Q^2$}
  \DashLine(130,0)(130,40){2}
  \DashLine(290,0)(290,40){2}
  \DashLine(170,15)(170,40){2}
  \LongArrow(172,18)(288,18)
  \LongArrow(288,18)(172,18)
  \LongArrow(132,3)(288,3)
  \LongArrow(288,3)(132,3)
  \Text(230,26)[]{\large $\ln 1/(x \kTpot{2}/Q^2)$}
  \Text(210,9)[]{\large $\ln 1/x$}
\end{picture}}}

  \caption{\label{fig:hard-probe} The hard colour-neutral probe
    $Q^2$ probes the gluon. The continuous line and the long-dashed
    line chains correspond to a struck gluon of $\kTpot{2} < Q^2$ and
    $\kTpot{2} > Q^2$, respectively.}}

We first study the relative importance of the different terms
contributing to the LDC integrated gluon distribution function. In
figure \ref{fig:int-terms} the integrated gluon distribution function is
shown as a function of $x$, for fixed $Q^{2} = 16 \, GeV^{2}$.  As can
be seen, at small $x$-values the first term of
eq.~(\ref{eq:int-terms}) dominates; as $x$ decreases the second term 
becomes noticeable. At large $x$-values, the behavior is governed 
by the input distribution function. 

\FIGURE[t]{\epsfig{file=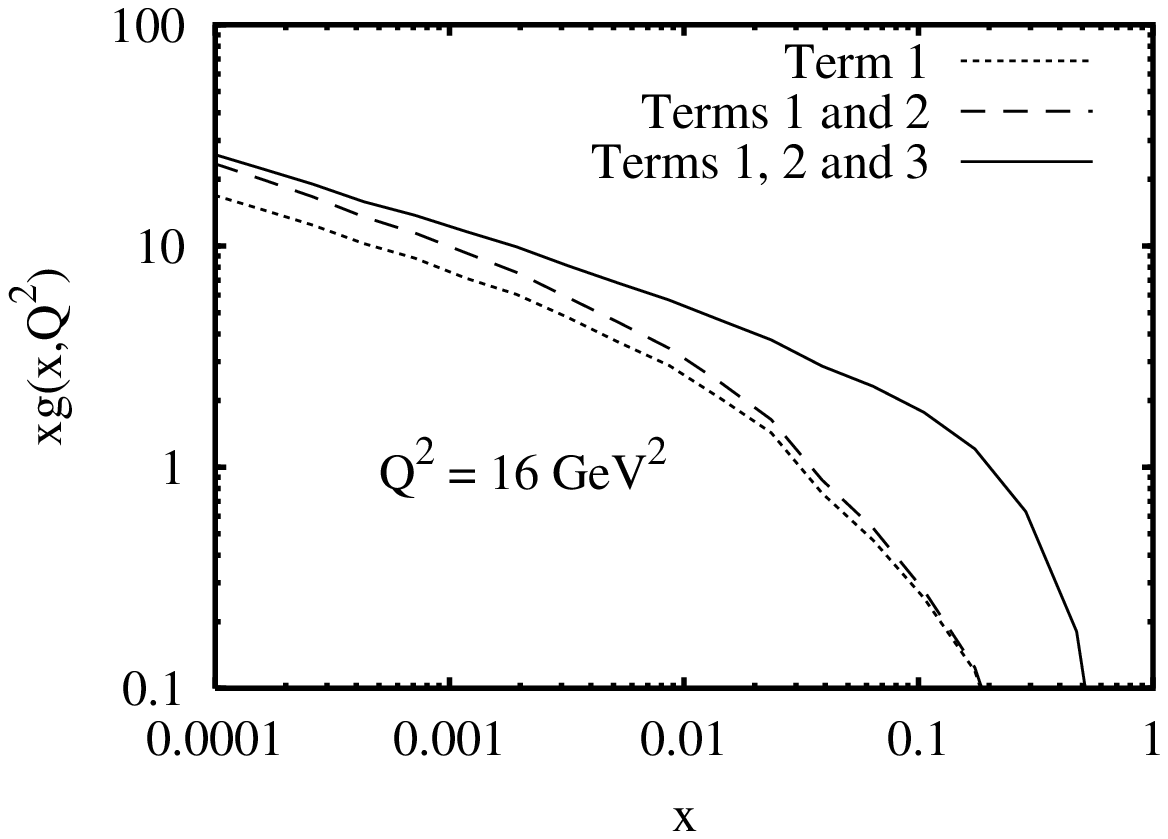,width=10cm}
  \caption{\label{fig:int-terms} Relative importance of the
    different terms contributing to the LDC integrated gluon distribution
    function, for fixed $Q^{2} = 16 \, GeV^{2}$. The dotted curve
    corresponds to keeping only the first term in
    eq.~(\ref{eq:int-terms}); keeping the first and second terms we
    obtain the dashed curve.  The total result is represented by the
    full curve.}}

Next we compare the LDC results with those obtained by other analyses.
In figure \ref{fig:int-comp} we show the LDC integrated gluon
distribution function as a function of $x$ for $Q^2 = 16\,GeV^2$ and
$Q^2 = 100\,GeV^2$. We show our result for the \textit{gluonic}
together with the \textit{gluonic-2} case. As discussed above, they
both give a good fit to $F_2$.  We see that the \ldcmc results lie
significantly below the JS curve for large $x$, but above JS for
smaller $x$-values.
\FIGURE[t]{\epsfig{file=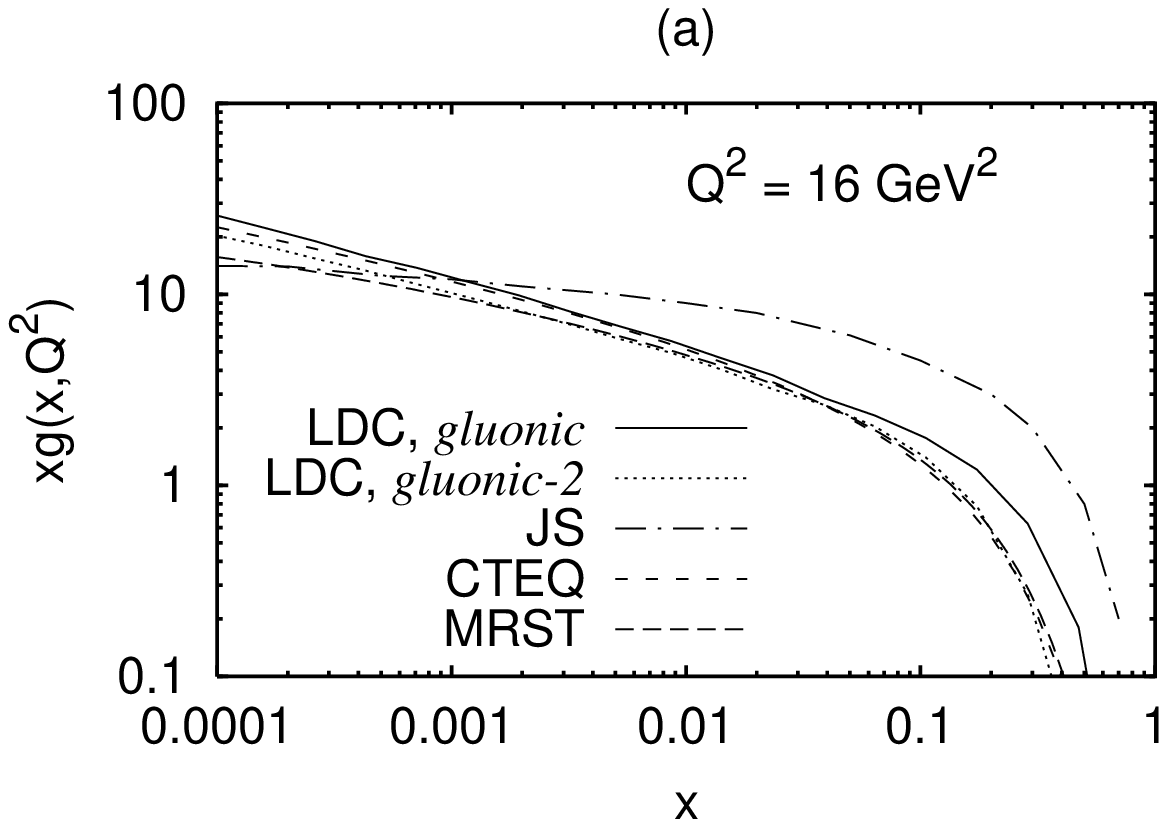,width=7.4cm}
  \epsfig{file=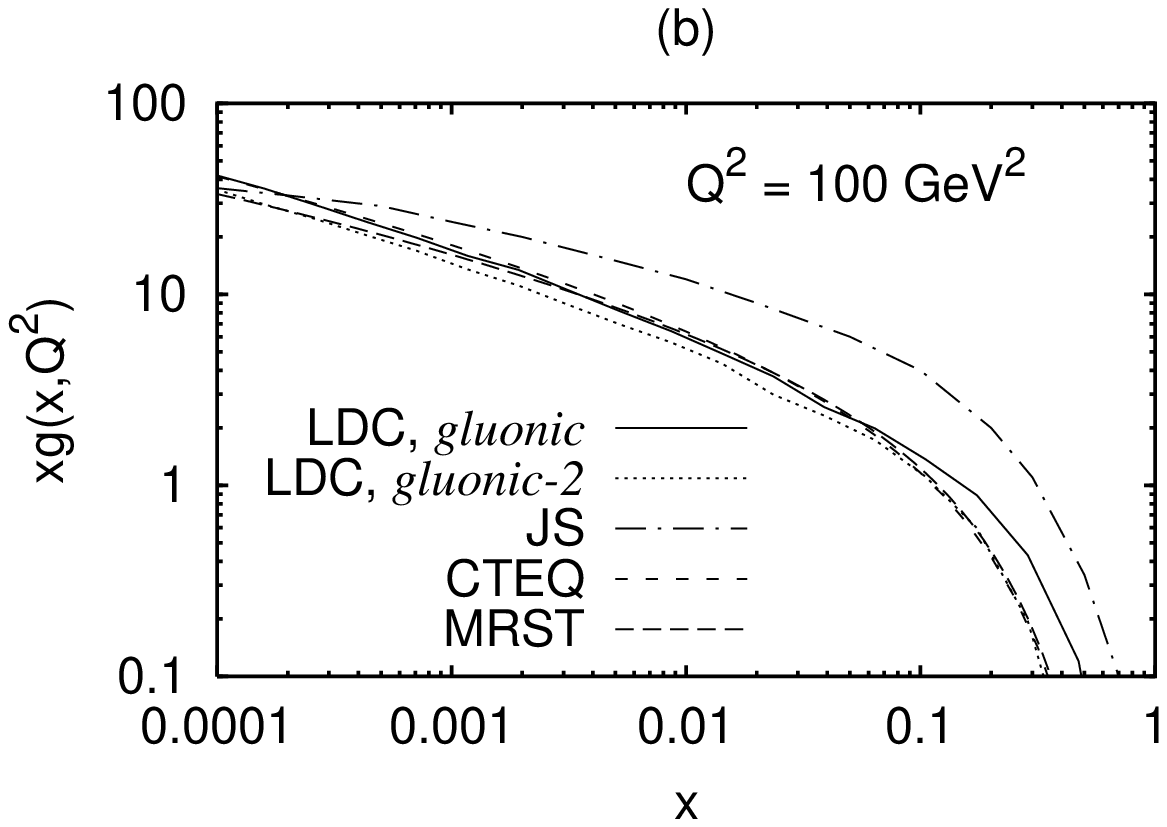,width=7.4cm}
  \caption{\label{fig:int-comp} The LDC integrated gluon
    distribution function (full curve is \textit{gluonic}, dotted curve
    is \textit{gluonic-2}), compared to the corresponding results of
    JS (dash-dotted curve), CTEQ (short-dashed curve) and MRST
    (long-dashed curve), for (a) $Q^2 = 16\,GeV^2$ and (b) $Q^2 =
    100\,GeV^2$.}}
Also shown in figure \ref{fig:int-comp} are the corresponding results
for CTEQ5M1 \cite{Lai:1999wy} and MRST20011 \cite{Martin:2001es}.  An
important point is that, while the LDC and JS results have been fitted
to $F_2$ data only, the CTEQ and MRST curves have been fitted to more
data. We see that these latter curves are more or less in agreement at
large $x$, where there is more data available, while they separate
more for smaller $x$-values.  Clearly the LDC result agrees well with
these curves for small $x$, where it almost coincides with CTEQ5M1.
For larger $x$ the LDC curve lies above CTEQ and MRST for
\textit{gluonic}, but agrees well with them for \textit{gluonic-2}.

This comparison leads to increased confidence in the physical
relevance of the LDC model in general and in the LDC unintegrated
gluon distribution function in particular. We shall study this in the
next subsection.

\subsection{Results for the Unintegrated Gluon Distribution Function}
\label{subsect-unintegrated}

We now turn our attention to the topic of unintegrated gluon distribution
functions, and we first want to study the effects from sub-leading
corrections caused by the inclusion of quarks and by the inclusion of
the non-singular terms in the gluon splitting function, in order to
verify the statements made earlier in this section.

In figure~\ref{fig:unint-quarks-Pgg} the unintegrated LDC gluon
distribution function is shown both as a function of $x$ and $\kTpot{2}$.
Comparing the result for the purely gluonic case to that when we allow
for quarks as well, we see that the differences are very small.  (Note
that the input distribution functions have been refitted.)
\FIGURE[t]{\epsfig{file=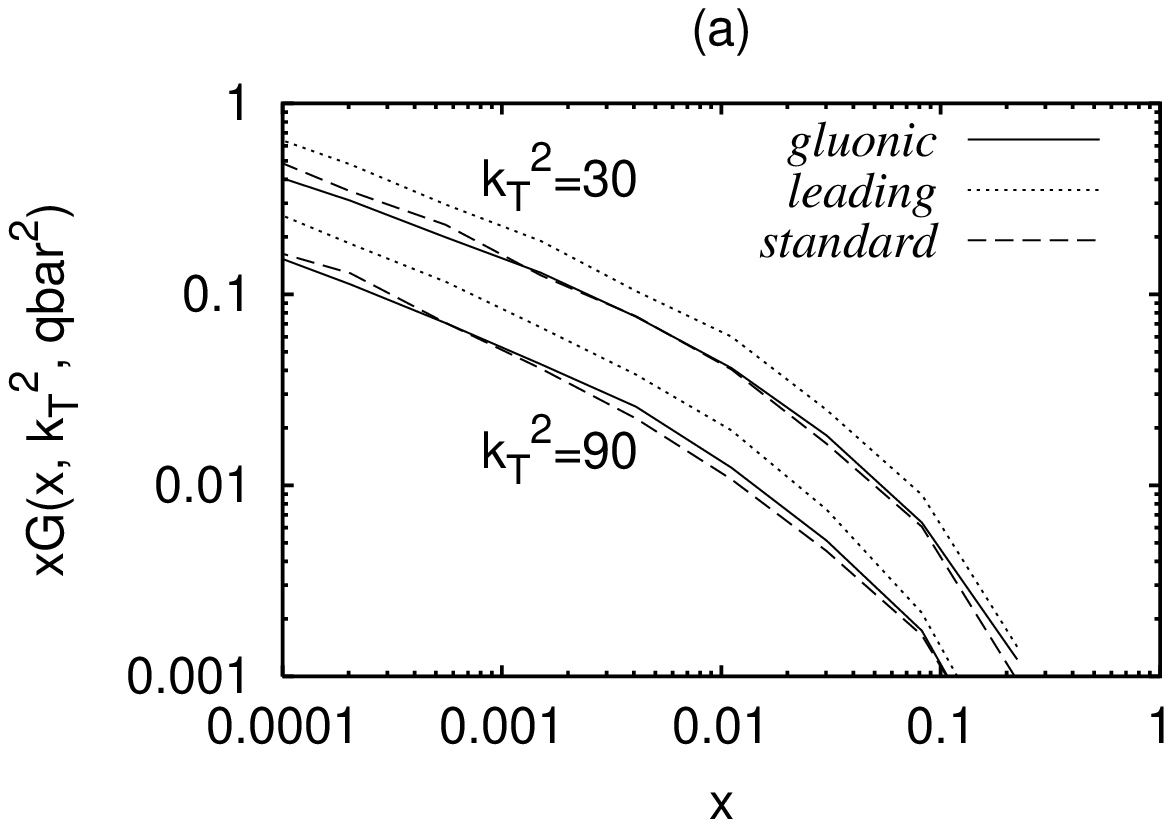,width=7.4cm}
  \epsfig{file=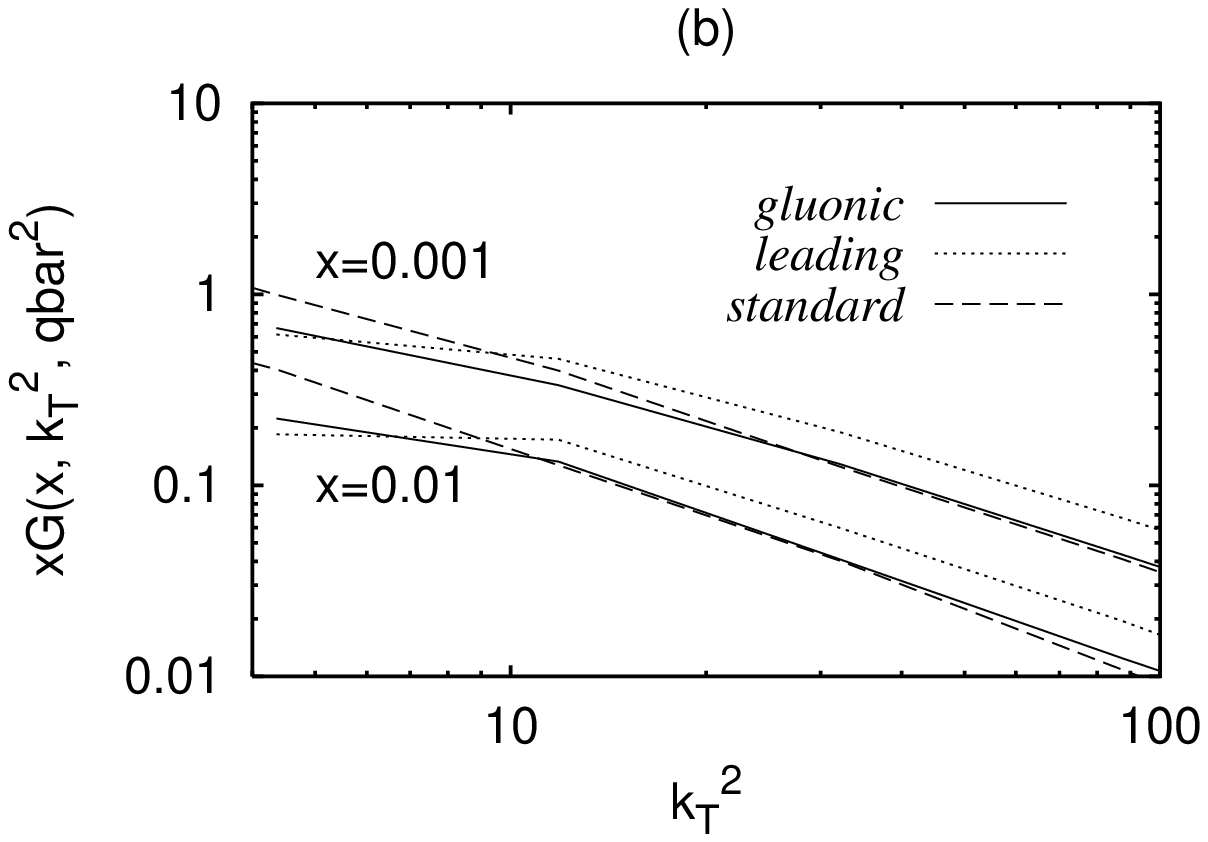,width=7.4cm}
  \caption{\label{fig:unint-quarks-Pgg}
    The LDC unintegrated gluon distribution function as a function of (a)
    $x$ and (b) $\kTpot{2}$. Study of sub-leading effects caused by
    inclusion of quarks and non-singular terms in the gluon splitting
    function. Full line is \textit{gluonic}, dotted line is
    \textit{leading} and dashed line is \textit{standard}. (In these 
    figures $\qbar$ is put equal to $\kT$.)}}
Even though the effect of omitting the non-singular terms in the gluon
splitting function is rather small, it is nevertheless not completely
negligible: as can be seen in figure~\ref{fig:unint-quarks-Pgg}a, at
small $x$ there is a discrepancy of about a factor of two, a result
that is related to the problem with the description of H1 forward-jet
data that was mentioned in Section \ref{sect-DIS}.

Our conclusion is that the non-leading effects can be largely
compensated by slight modifications of the input distribution
functions; we note in particular that the results with and without 
quark links are almost identical.

\FIGURE[t]{\epsfig{file=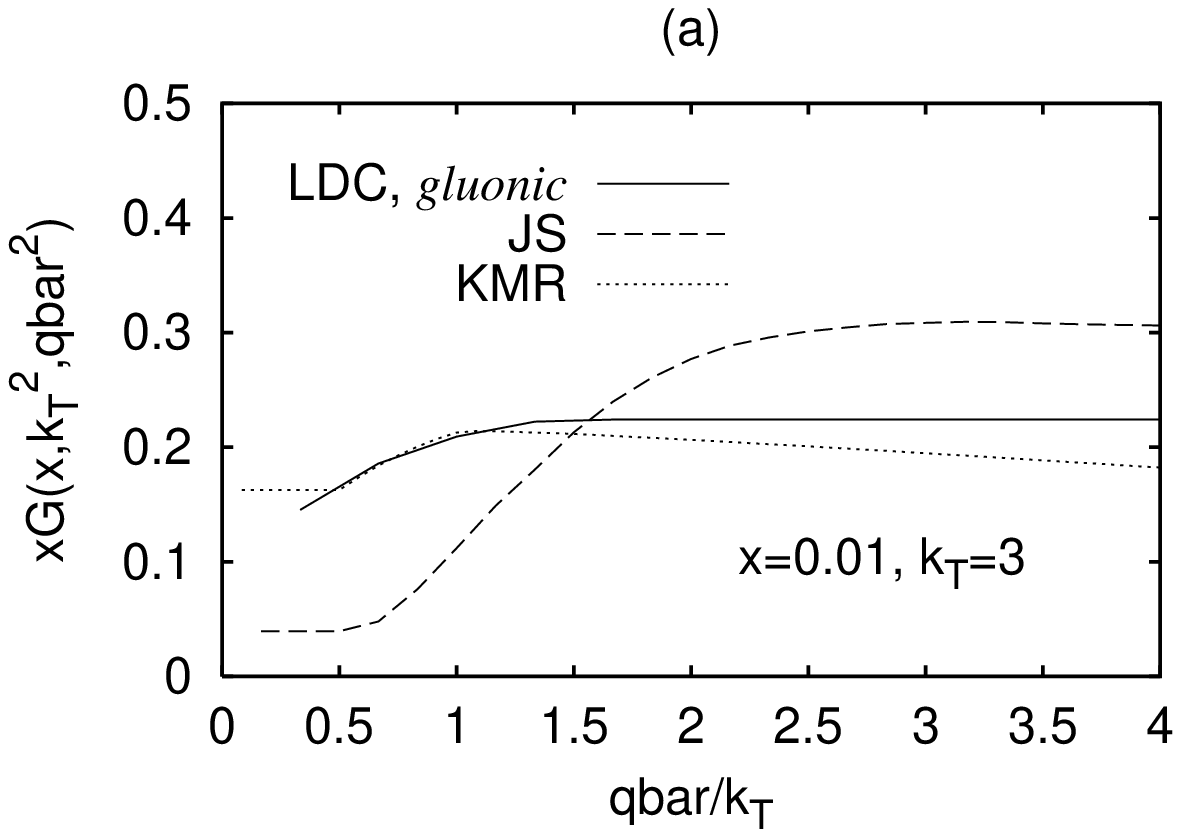,width=7.4cm}
  \epsfig{file=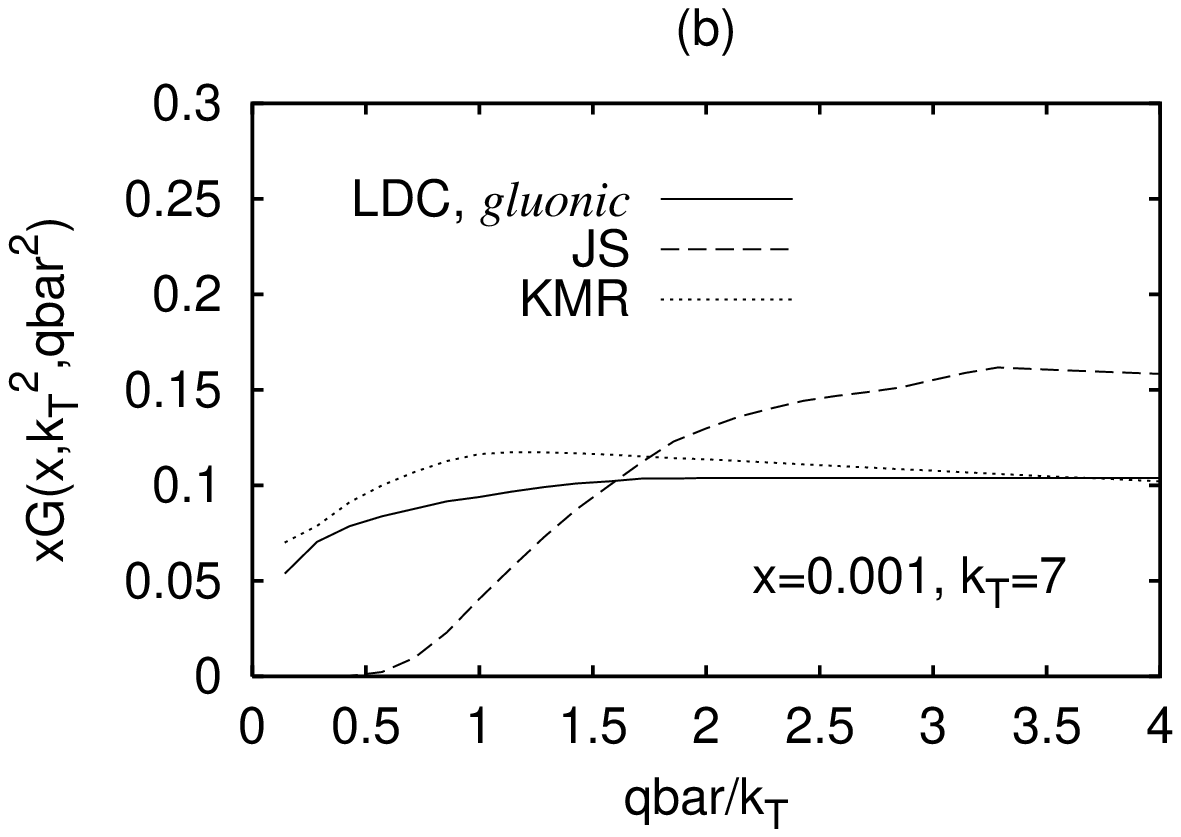,width=7.4cm}
  \caption{\label{fig:unint-comp} The LDC \textit{gluonic} unintegrated
    gluon distribution function (full curve), compared to the
    corresponding results of JS (long-dashed curve) and KMR (dotted
    curve), as functions of $\qbar/\kT$ for (a) $x=0.01$ and $\kT=3 GeV$
    and (b) $x=0.001$ and $\kT=7 GeV$.}}

We now want to study the dependence upon the two different scales
discussed in section \ref{sect-different}. We show in figure
\ref{fig:unint-comp} the dependence on $\qbar$ for fixed $\kT$ for the
LDC, CCFM, and KMR formalisms. As discussed in section \ref{sect-different},
the $\qbar$-dependence is rather weak in the LDC model, but very
strong in the CCFM approach. We also see that the result in the KMR
formalism is quite insensitive to variations in $\qbar$. This is also
illustrated in figure \ref{fig:unint-comp-2sc}, which shows the
results as functions of $\kT^2$ for fixed $\qbar$ = 10~GeV.  We see
that the JS results start to fall dramatically below the other two, as
$\kT$ approaches $\qbar$.

\FIGURE[t]{\epsfig{file=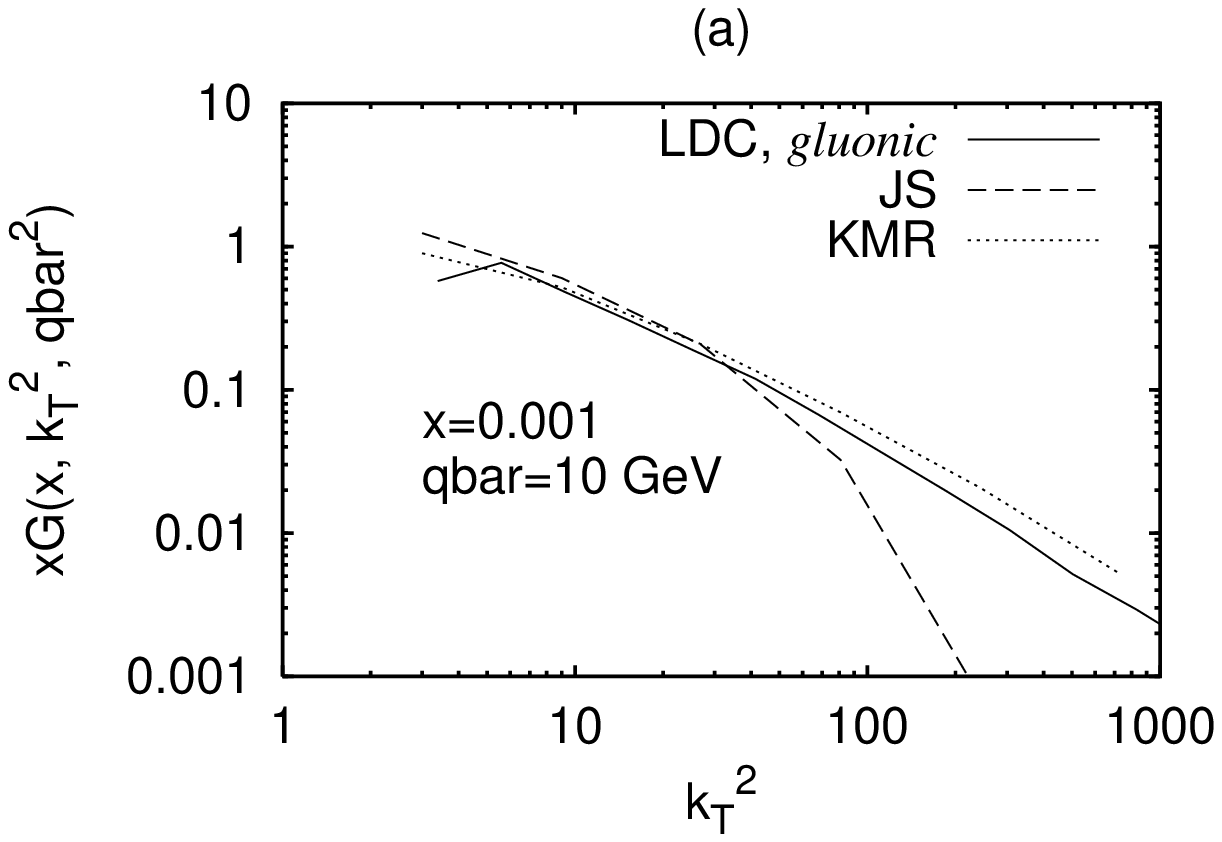, width=7.4cm}
  \epsfig{file=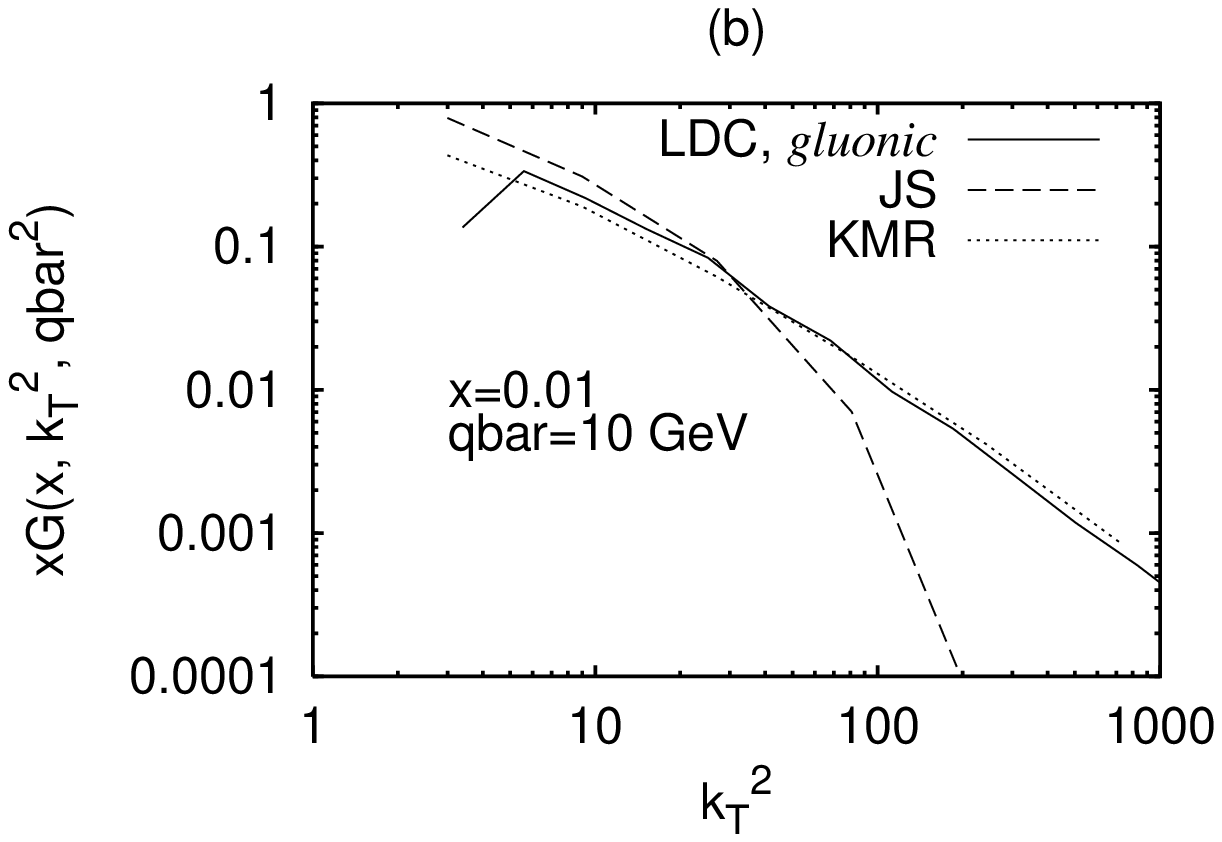,width=7.4cm}
  \caption{\label{fig:unint-comp-2sc} The LDC \textit{gluonic}
    unintegrated gluon distribution function (full curve), compared to
    the corresponding results of JS (long-dashed curve) and KMR
    (dotted curve), as functions of $\kTpot{2}$ for (a) $x=0.001$ and
    (b) $x=0.01$. (In these figures $\qbar$ is put equal to $10 \, GeV$.)}}

From figure \ref{fig:unint-comp} we note, however, that the CCFM
result saturates for $\qbar \gtrsim 2 \kT$. In a hard-interaction
event the scales $\hat{|t|}$ and $\hat{s}$ are normally larger than
$\kT^2$, and often characteristically by a factor of this order.  For
this reason we want to argue that when comparing the different
formalisms, it is more relevant to study the CCFM distributions for
$\qbar$ equal to the saturation value, rather than e.g.\ for $\qbar =
\kT$. Indeed, for this larger $\qbar$-value we see in figure
\ref{fig:unint-comp} that there is a rather good agreement between the
three models.

This feature is further illustrated in figure
\ref{fig:unint-comp-1sc}, which shows the distribution functions for
$\qbar = 2 \kT$, as a functions of $\kT^2$ for fixed $x$ and as a
functions $x$ for fixed $\kT$, and we see indeed a reasonable
agreement between the LDC, JS, and KMR results. In these figures we
also show the single scale KMS and dGRV results. Although these earlier
parameterizations are somewhat lower for larger $x$-values, we note a
fair overall agreement between all five models.

\FIGURE[t]{\epsfig{file=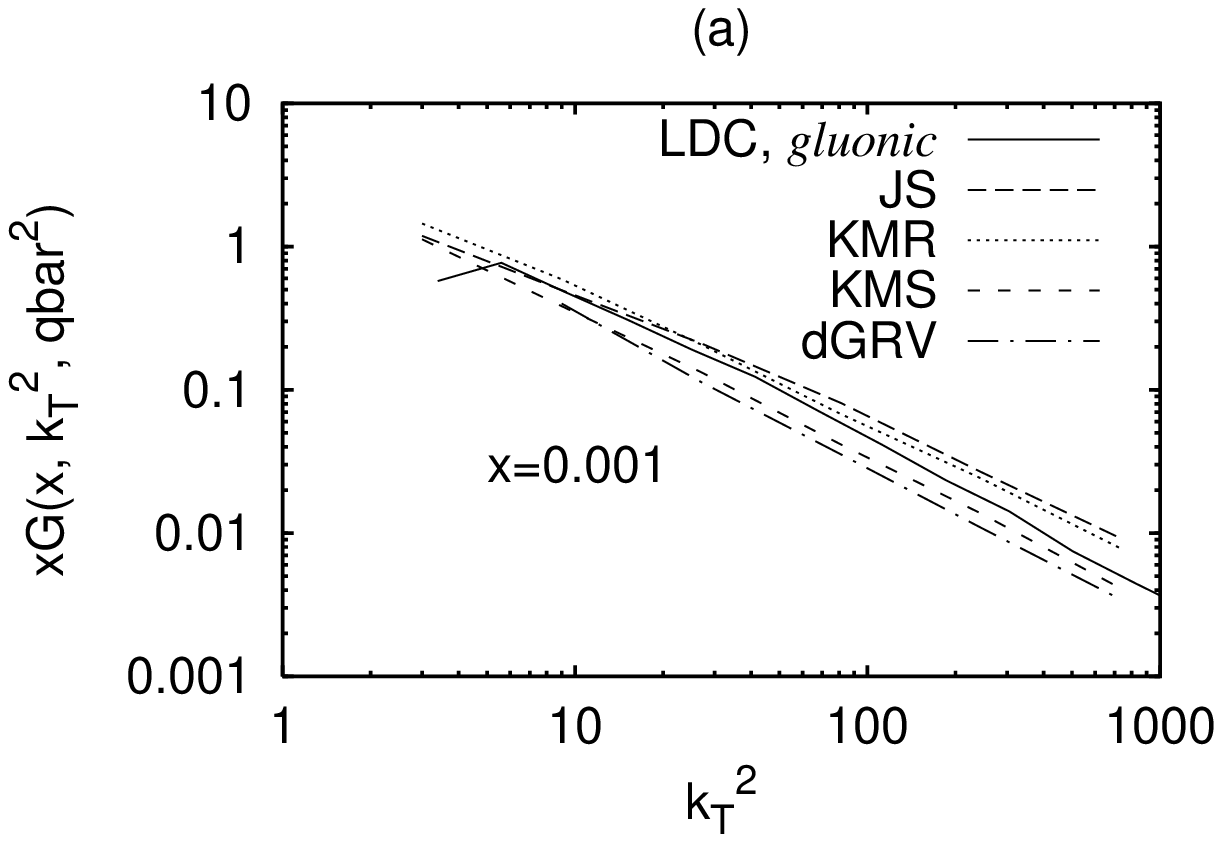,width=7.45cm}
  \epsfig{file=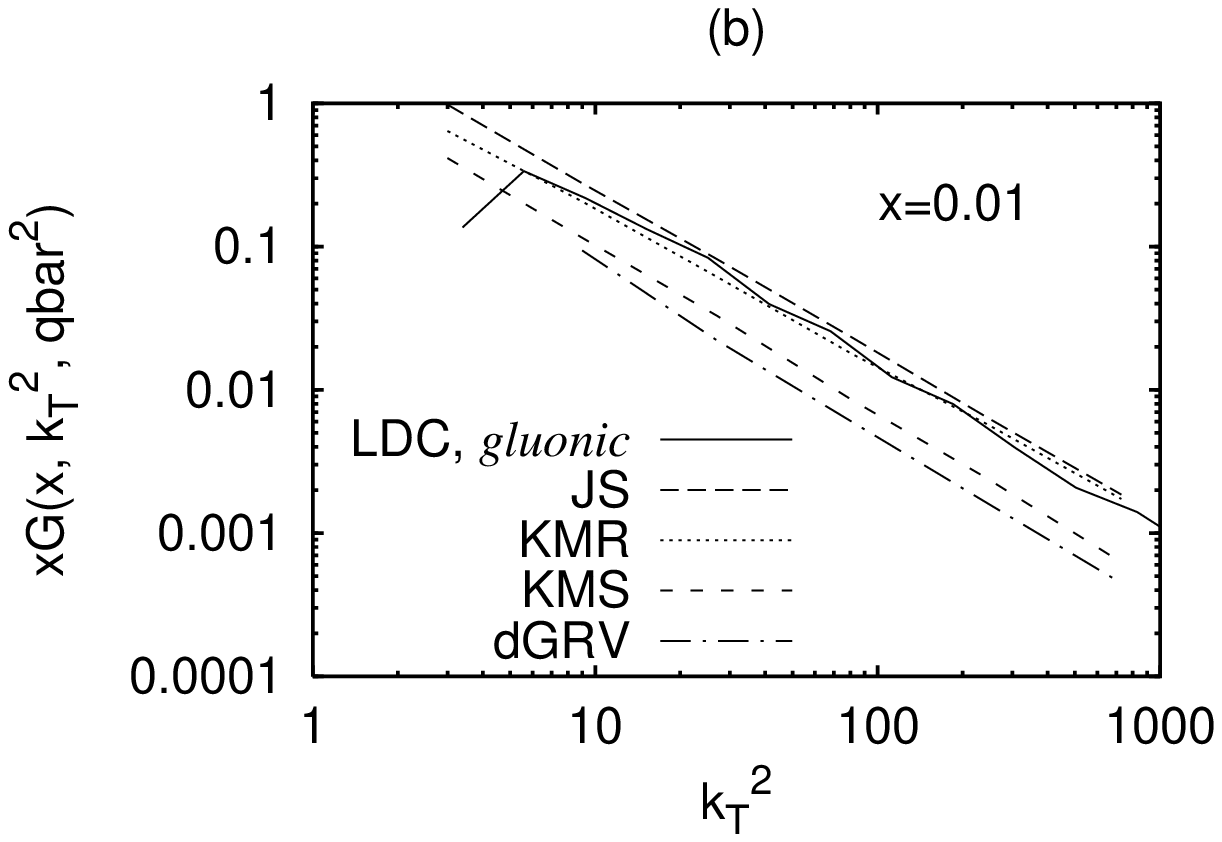,width=7.45cm}
  \epsfig{file=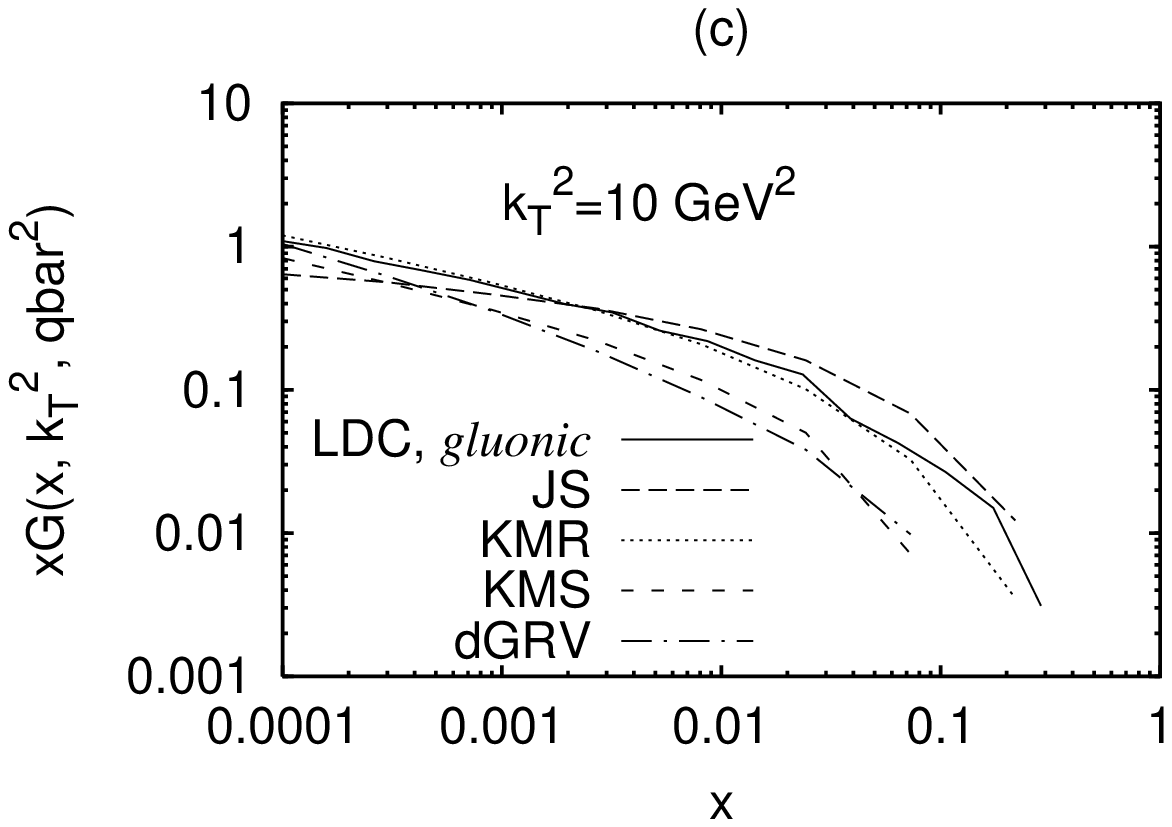,width=7.45cm}
  \epsfig{file=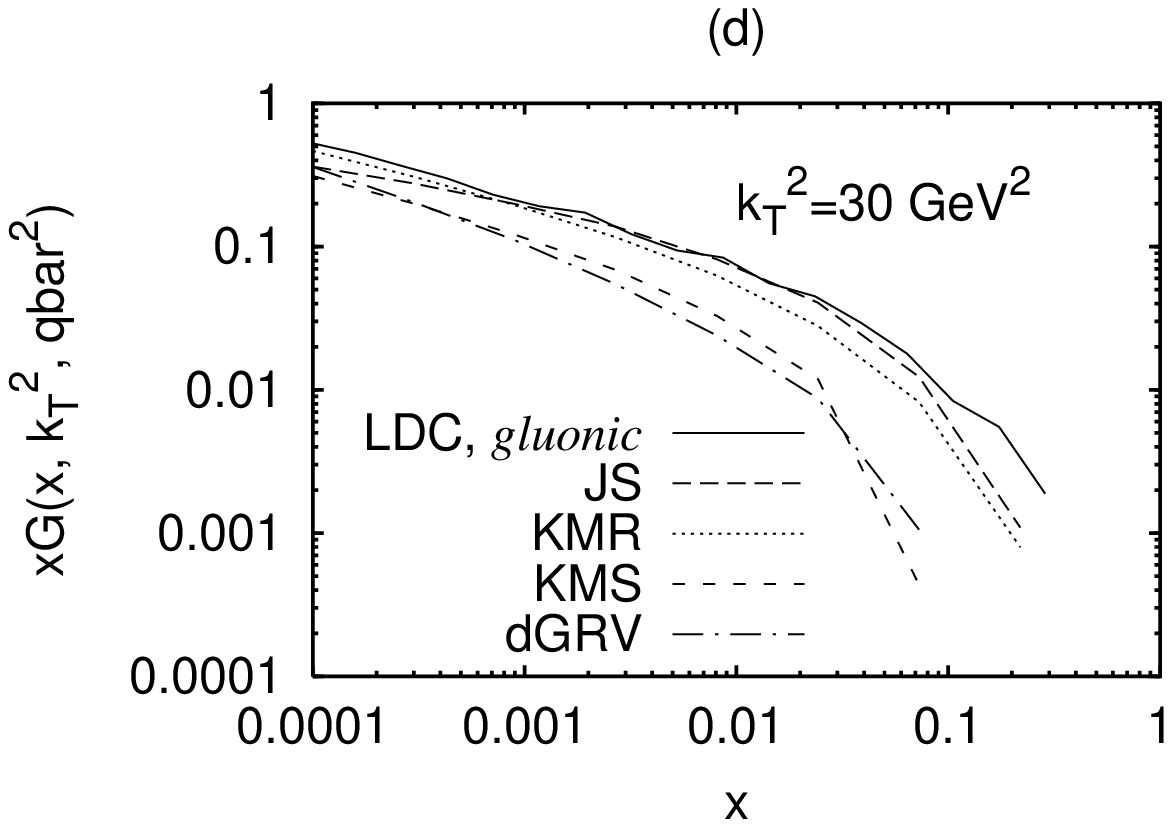,width=7.45cm}
  \caption{\label{fig:unint-comp-1sc} The LDC \textit{gluonic}
    unintegrated gluon distribution function (full curve), compared to
    the corresponding results of JS (long-dashed curve), KMR (dotted
    curve), KMS (short-dashed curve) and dGRV (dash-dotted curve)
    as functions of $\kTpot{2}$ for (a) $x=0.001$ and (b) $x=0.01$,
    and as functions of $x$ for (c) $\kTpot{2}= 10\,GeV^2$ and (d)
    $\kTpot{2}= 30\,GeV^2$. Results for LDC, JS and KMR, with 
    $\qbar=2 \kT$, are shown together with the 1-scaled distribution
    functions of KMS and dGRV.}}
%

\section{Summary}
\label{sect-summary}

Unintegrated parton distribution functions are not uniquely defined.
They are not experimental observables, and their definition and
properties depend critically on the formalism used. Thus when
calculating e.g.\ cross-sections for production of jets or heavy
quarks it is necessary to use a consistent formalism for off-shell
parton cross-sections and parton distribution functions.
 
In this paper we present results for integrated and unintegrated gluon
distribution functions according to the definitions in the Linked
Dipole Chain model, obtained from the \ldcmc program. We compare them
with those obtained in other formalisms, in particular with results
from the CCFM model obtained from the \smallx and \cascade MCs, and we
demonstrate how to make a relevant comparison between the models.
Indeed we find in this way a reasonable agreement between
distributions obtained in different formalisms.
 
Adjusting the input distribution functions for $\kT = k_{\perp0}$, it is
possible to find a good fit to the structure function $F_2$ from
\ldcmc. The corresponding integrated gluon distribution function agrees
well with fits to more complete data sets obtained by the CTEQ or MRST
collaborations. The result is rather insensitive to non-singular
contributions in the gluon splitting function or from from quark links
in the chain. The contributions from quarks can be almost fully
compensated by adjusting the input distribution functions. Omitting the
non-singular terms in the gluon splitting function has a somewhat
larger effect, reducing the gluon distribution for small $x$ by
roughly a factor 2.
 
In the CCFM formalism the unintegrated parton distribution functions
depend sensitively on two different scales: $\kT$, which specifies the
transverse momentum and the virtuality of the interacting parton, and
$\qbar$, which determines an angle beyond which there is no (quasi-)real 
parton in the chain of initial-state radiation. Many of the
gluons which make up the initial-radiation chain in the CCFM model
are treated as final-state radiation in the LDC formalism. Therefore
typical $z$-values are smaller in LDC, and most of the problem of angular
ordering is postponed to the treatment of the final-state radiation.
This implies that in the LDC formalism the gluon distribution function
is quite insensitive to the second scale $\qbar$, and to leading $\log
1/x$ it depends only on a single scale $\kT$.
 
In a typical hard sub-collision the relevant additional scale $\qbar$
should be given by $\qbar^2\sim|\hat{t}|$ or $\hat{s}$. We
observe that in the CCFM formalism, for a fixed $\kT$, the gluon
distribution function saturates for $\qbar \gtrsim 2 \kT$. This
corresponds to typical values of $\hat{t}$ or $\hat{s}$, and we
therefore suggest that for a relevant comparison between the different
formalisms we should choose $\qbar$ in this saturation region. Here we
find indeed a reasonable agreement, not only between LDC and CCFM, but
also with e.g.\ the two-scale formalism presented by Kimber, Martin
and Ryskin, and the one-scale formalism by Kwiecinski, Martin and
Sutton. Actually we see that it also agrees rather well with the
derivative of the integrated distribution function GRV98, obtained in
the pure DGLAP formalism.

\section*{Acknowledgments}

We want to thank Hannes Jung for valuable discussions, and in
particular for help in evaluating the distribution functions for the JS,
KMS, and KMR models.

\bibliographystyle{utcaps}  
\bibliography{references} 

\end{document}